\documentclass[twocolumn,showpacs,preprintnumbers,amsmath,amssymb]{revtex4}
\usepackage{graphicx}% Include figure files
\usepackage{dcolumn}% Align table columns on decimal point
\usepackage{bm}% bold math
\newif\ifappendixon
\begin{document}

\preprint{APS/123-QED}

\title{Slow-light solitons revisited}

\author{A.V. Rybin\footnote{http://www.slowlight.org} }
\affiliation{Department of Physics, University of Jyv\"askyl\"a PO
Box 35, FIN-40351
\\ Jyv{\"a}skyl{\"a}, Finland\\ St Petersburg University of Information Technologies,
Mechanics and Optics, Kronwerkskii ave 49, 197101, St Petersburg,
Russia } \email{andrei.rybin@phys.jyu.fi}

\author{ I.P. Vadeiko}
\affiliation{Physics Department, McGill University 3600 rue
University Montreal, QC H3A 2T8, Canada } \email{vadeiko@gmail.com}

\author{ A. R. Bishop}
\affiliation{Theoretical Division and Center for Nonlinear Studies,
Los Alamos National Laboratory, Los Alamos, New Mexico 87545, USA}
\email{arb@lanl.gov}

\date{August 11, 2006}

\begin{abstract}
We investigate  propagation of  slow-light solitons in atomic media
described by the nonlinear $\Lambda$-model. Under a physical
assumption, appropriate to the slow light propagation, we reduce the
$\Lambda$-scheme to a simplified nonlinear model, which is also
relevant to 2D dilatonic gravity. Exact solutions describing various
regimes of stopping slow-light solitons can then be readily derived.

\end{abstract}
%Use showkeys class option if keyword
\maketitle

Recent progress in experimental techniques for the coherent control
of light-matter interaction opens many opportunities for interesting
practical applications. The experiments are carried out on various
types of materials such as cold sodium atoms \cite{Hau:1999,
Liu:2001}, rubidium atom vapors \cite{Phillips:2001, Bajcsy:2003,
Braje:2003, Mikhailov:2004}, solids \cite{Turukhin:2002,
Bigelow:2003}, and photonic crystals \cite{Soljacic:2004}. These
experiments are based on the control over the absorption properties
of the medium and study slow light and superluminal light effects.
The control can be realized in the regime of electromagnetically
induced transparency (EIT), by the coherent population oscillations
or other induced transparency techniques. The use  of each different
material brings specific advantages important for the practical
realization of the effects. For instance, the cold atoms have
negligible Doppler broadening and small collision rates, which
increases ground-state coherence time. The experiments on rubidium
vapors are carried at room temperatures and this does not require
application of complicated cooling methods. The solids are a strong
candidate for realization of long-living optical memory. Photonic
crystals provide a broad range of paths to guide and manipulate slow
light. The interest in the physics of light propagation in atomic
vapors and Bose-Einstein condensates (BEC) is strongly motivated by
the success of research on storage and retrieval of optical
information in these media \cite{Hau:1999, Liu:2001, Phillips:2001,
Kocharovskaya:2001, Bajcsy:2003, Dutton:2004}.

Even though the linear approach to describing these effects based on
the theory of electromagnetically induced transparency (EIT)
\cite{Harris:1997} is developed in detail \cite{Lukin:2003}, modern
experiments require more complete nonlinear descriptions
\cite{Dutton:2004}. The linear theory of EIT assumes the probe field
to be  much weaker than the control field. To allow significant
changes in the initial atomic state due to interaction with the
optical pulse, in our consideration we go beyond the limits of
linear theory. In the adiabatic regime, when the fields change in
time very slowly, approximate analytical solutions \cite{Grobe:1994,
Eberly:1995} and self-consistent solutions \cite{andreev:1998} were
found and later applied in the study of processes of storage and
retrieval \cite{Dey:2003}. Different EIT and self-induced
transparency solitons in nonlinear regime were classified and
numerically studied for their stability \cite{Kozlov:2000}. As  was
demonstrated by Dutton and coauthors \cite{Dutton:2001} strong
nonlinearity can result in interesting new phenomena. Recent
experiments and numerical studies \cite{Matsko:2001, Dutton:2004}
have shown that the adiabatic condition can be relaxed, allowing for
much more efficient control over the storage and retrieval of
optical information.

\begin{figure}
\includegraphics[width=60mm]{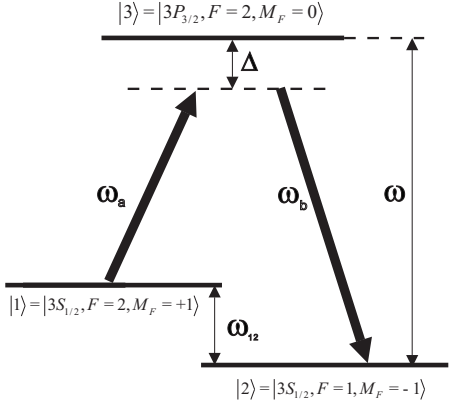}
\caption{\label{fig:spec1} The  $\Lambda$-scheme for working energy
levels of sodium atoms. { The parameters of the scheme are:
$\omega_{12}/(2\pi)=1772 \mathrm{MHz}$, $\omega/(2\pi)=5.1\cdot
10^{14} \mathrm{Hz}$ ($\lambda=589 \mathrm{nm}$), and $\Delta$ is
the variable detuning from the resonance.}}
\end{figure}

In this paper we study the interaction of light with a gaseous
active medium whose working energy levels are well approximated by
the $\Lambda$-scheme. Our theoretical model is a very close
prototype for a gas of sodium atoms, whose interaction with the
light is approximated by the structure of levels of the
$\Lambda$-type. The structure of levels is given in
Fig.~\ref{fig:spec1}, where two hyperfine sub-levels of sodium state
$3^2S_{1/2}$ with $F=1, F=2$ are associated with $|2\rangle$ and
$|1\rangle$ states, {respectively~\cite{Hau:1999}}. The excited
state $|3\rangle$ corresponds to the hyperfine sub-level of the term
$3^2P_{3/2}$ with $F=2$. We consider the case when the atoms are
cooled down to microkelvin temperatures in order to suppress the
Doppler shift and increase the coherence life-time for the ground
levels. The atomic coherence life-time in sodium atoms at a
temperature $0.9 {\mu}$K is of the order of 0.9 ms \cite{Liu:2001}.
Typically, in the experiments the pulses have a length of
microseconds, which is much shorter than the coherence life-time and
longer than the optical relaxation time of $16.3 ns$.

The gas cell is illuminated by two circularly polarized optical
beams co-propagating in the z-direction. One beam, denoted as
channel $a$, is a $\sigma^-$-polarized field, and the other,
denoted as $b$, is a $\sigma^+$-polarized field. The corresponding
fields are presented within the slow-light varying amplitude and
phase approximation (SVEPA) as
\begin{equation}\label{fields}
     \vec{E}=\vec{e}_a\, \mathcal{E}_a e^{i(k_a z-\omega_a t)} +
     \vec{e}_b\, \mathcal{E}_b e^{i(k_b z-\omega_b t)} +c.c.
\end{equation}
Here, $k_{a,b}$ are the wave numbers, while the vectors
$\vec{e}_a,\vec{e}_b$ describe polarizations of the fields. It is
convenient to introduce two corresponding Rabi frequencies:
\begin{equation}\label{Rabi_f}
    \Omega_a=\frac{2\mu_{a}\mathcal{E}_a}{\hbar},
    \Omega_b=\frac{2\mu_{b}\mathcal{E}_b}{\hbar},
\end{equation}
where $\mu_{a,b}$ are dipole moments of quantum transitions in the
channels $a$ and $b$.

Within the SVEPA, and in the sharp line limit case ($\Delta=0$), the
wave equations for two  Rabi-frequencies are reduced to the first
order PDEs:
\begin{eqnarray}\label{Maxwell_r}
    \partial_\zeta \Omega_a = i \nu_0 \,\psi_3 \psi_1^*, \;
    \partial_\zeta \Omega_b = i \nu_0 \,\psi_3 \psi_2^*.\quad
\end{eqnarray}
Here $\zeta=z/c, \tau=t-z/c$, and $\nu_0$ is a coupling constant,
which depends on the density of atoms.

The Schr\"odinger equation for the amplitudes $\psi_{1,2,3}$ of
atomic wave function reads
\begin{eqnarray}\label{Schrod}
    \partial_\tau \psi_1 &=& \frac i2 \Omega_a^*\,\psi_3;\nonumber\\
    \partial_\tau \psi_2 &=& \frac i2 \Omega_b^*\,\psi_3;\\
    \partial_\tau \psi_3 &=& -i\gamma\,\psi_3 +
    \frac i2(\Omega_a\psi_1+\Omega_b \psi_2).\nonumber
\end{eqnarray}
Here %$\Delta$ is the variable detuning from the resonance,
$\gamma$
describes the relaxation rate, and we set
 $\hbar=1$.
 We can now exclude the amplitudes of the lower levels $\psi_{1,2}$
 and rewrite Eqs.(\ref{Maxwell_r}), (\ref{Schrod}) in the
form:
\begin{eqnarray}\label{MB_eqs}
   & \frac1{\psi_3^*}\partial_{\tau}\frac1{\psi_3}\partial_{\zeta}\Omega_a
  = \frac{\nu_0}{2}\Omega_a;\nonumber\\
   & \frac1{\psi_3^*}\partial_{\tau}\frac1{\psi_3}\partial_{\zeta}\Omega_b
  = \frac{\nu_0}{2}\Omega_b;\\
   & \partial_\tau |\psi_3|^2 =-\gamma\,|\psi_3|^2  -\frac1{2\nu_0}\partial_{\zeta}
    (|\Omega_a|^2+|\Omega_b|^2).\nonumber\\
   & \partial_\tau \varphi_3 =   -\frac1{2\nu_0\,|\psi_3|^2}
   (|\Omega_a|^2\,\partial_{\zeta}\varphi_a+
   |\Omega_b|^2\,\partial_{\zeta}\varphi_b).\nonumber
\end{eqnarray}
Here, $\varphi_{a,b,3}$ are the phases of the fields $\Omega_{a,b}$
and $\psi_3$, respectively. For simplicity, and without a loss of
generality, we assume in Eqs.(\ref{MB_eqs}) that the fields
$\Omega_{a,b}$ are real, i.e $\varphi_{a,b}=0$. Therefore, we can
choose $\varphi_3=0$. Notice that the first two equations are wave
equations for the fields in curvilinear space described by the
metric depending on the amplitude of the excited state $\psi_3$.

To make parameters dimensionless, we measure the time in units of
the optical pulse length $t_p=1\mu s$ typical for the experiments on
the slow-light phenomena~\cite{Hau:1999}. Therefore, the Rabi
frequencies are normalized to $\mathrm{MH}z$. The spatial coordinate
will be normalized to the spatial length of the pulse slowed down in
the medium to several meters per second. According to the linear
theory, the group velocity of slow-light pulse is $v_g \approx c
\frac {\Omega_0^2}{2\nu_0}$. Here $\Omega_0$ is a magnitude of the
controlling field required in EIT experiments. Typically, this field
has a magnitude of order of few megahertz. So, we choose
$\Omega_0=3$ and $v_g =10^{-7} c$ as representative values reported
in experiments with BEC of sodium atoms. Hence the pulse spatial
length is $l_p=v_g t_p=30 \mu m$, and $\zeta$ is normalized to
$10^{-13} s$. In the dimensionless units, the coupling constant
$\nu_0=\frac {\Omega_0^2}2=4.5$.

In the absence of relaxation, i.e. for $\gamma=0$, the system of
equations Eqs.(\ref{Maxwell_r}),(\ref{Schrod}) is exactly solvable
in the framework of the inverse scattering  method (IS)
\cite{fad,Park:1998,gab,Rybin:2004}. In the present work we provide
an elementary method to derive slow-light solitons for the case of
an arbitrary controlling field.

In the context of slow light phenomena, the system is assumed to be
initially in the following stationary state:

\begin{equation}\label{init_fields0}
    \Omega^{}_a=0,\; \Omega^{}_b=\Omega(\tau),\;
    |\psi_{at}\rangle=|1\rangle.
\end{equation}  Notice that the state $|1\rangle$
 is a dark-state for the controlling field $\Omega(\tau)$.
This means that the atoms do not interact with the field
$\Omega(\tau)$ created by the auxiliary laser. The configuration
Eq.(\ref{init_fields0}) above corresponds to a typical experimental
setup (see e.g. \cite{Hau:1999,Liu:2001,Bajcsy:2003}).

We intend to study the dynamics of coupled atom-field modulations in
the $\Lambda$-type model, which can preserve their spatial shape to
a large extent while propagating in the media. We consider such
solutions as a generalization of the dark-state polariton
\cite{Fleischhauer:2000}. In the linear theory the probe field only
appears in $\Omega_a$, whereas in the nonlinear theory it  also
forms an inseparable nonlinear superposition with the controlling
field in the channel $b$. However, in both cases the Rabi-frequency
$\Omega_a$ describes probe field modulations. Indeed, the field in
the channel $a$ induces atomic transitions from the state
$|1\rangle$ to the excited state $|3\rangle$. On the other hand, the
amplitude of the excited state drives the field $\Omega_a$. From the
results of linear and nonlinear theories of electromagnetically
induced transparency (EIT) \cite{Harris:1997, Grobe:1994, Dey:2003,
Dutton:2004, Rybin:2004, ryb12} we can infer a physically plausible
assumption that the population of the upper level is proportional to
the intensity of the field in the probe channel $a$, i.e.
$|\Omega_a|^2\sim |\psi_3|^2$. In the present work we assume that
this observation is relevant for slow light phenomena.
   Therefore, we postulate that

\begin{equation}
|\psi_3|^2=\frac{2}{\nu_0}k|\Omega_a|^2,\label{central}
\end{equation}
where  $k$ is an arbitrary parameter.

We emphasize that the imposed constraint   Eq.(\ref{central})
reduces Eqs.~(\ref{MB_eqs}) to a simplified nonlinear system, which
provides adequate descriptions of the slow light propagation. In
this sense the relation Eq.(\ref{central}) is {\it central} for the
present work. As we show below this condition is {\it sufficient}
and {\it necessary} for the slow-light solitons to exist.
Introducing new notations: $|\Omega_a|\equiv e^{-\rho},
\Omega_b=\eta$, we find from Eq.(\ref{MB_eqs}) together with
Eq.(\ref{central}) that the field $\rho$ satisfies the Liouville
equation

\begin{eqnarray}\label{Liouv}
    &\partial_{\zeta \tau}\rho=-k\,e^{-2\rho}.
\end{eqnarray}
together with the constraint
\begin{eqnarray}
%&\left(4k(\partial_\tau+\gamma)+\partial_\zeta\right)e^{-2\rho}+\partial_\zeta\eta^2=0,\label{dilaton_1}\\
&\partial_{\zeta \tau}\eta+\partial_\tau \rho\;\partial_\zeta \eta
=k\,e^{-2\rho}\,\eta, \label{dilaton_2}\end{eqnarray}

and an auxiliary equation
\begin{eqnarray}
&\left(4k(\partial_\tau+\gamma)+\partial_\zeta\right)e^{-2\rho}+\partial_\zeta\eta^2=0.\label{dilaton_1}%\\
%&\partial_{\zeta \tau}\eta+\partial_\tau \rho\;\partial_\zeta \eta
%=k\,e^{-2\rho}\,\eta.
%\label{dilaton_2}
\end{eqnarray}

It is interesting that the Liouville equation Eq.(\ref{Liouv})
appears in 2D gravity~\cite{Giacomini:2003} and describes a
gravitational field defined by the metric $g_{ab}=e^{-2\rho}
\gamma_{ab}$, where $\gamma_{ab}$ is the 2-dimensional Minkowski
metric. This connection to 2D gravity is further emphasized by the
observation that any solution of dilatonic
equations~\cite{Giacomini:2003} satisfies the system of slow-light
Eqs.(\ref{Liouv}), (\ref{dilaton_2}). The dilatonic equations have
the following form
\begin{equation}\label{GR}
\partial_{\zeta \tau}\rho-\frac{\partial_\eta V(\eta)}4\,
e^{-2\rho}=0,\;
\partial_{\zeta \tau}\eta+\frac{V(\eta)}2\,e^{-2\rho}=0,
\end{equation}
together with the constraint
\begin{equation}
\partial_{\zeta \tau}\eta+2\,\partial_\tau \rho\;\partial_\zeta
\eta=2k{\cal A}\,e^{-2\rho}\;.\label{dilaton_3}
\end{equation}
Here ${\cal A}(\zeta,\tau)$ is an arbitrary source term and
$V(\eta)=4k\left({{\cal A}-\eta}\right)$ plays the role of the
dilatonic potential. For the realization ${\cal
A}(\zeta,\tau)=\partial_\tau m(\tau)$ with an arbitrary function
$m(\tau)$,  the equations Eqs.(\ref{GR}),(\ref{dilaton_3}), and
Eq.(\ref{dilaton_1}) for $\gamma=0$ can be readily solved, viz.

\begin{eqnarray}\label{sol1}
&\rho=-\frac12 \log\left[{ \frac{\partial_\zeta\,A_+(\zeta)
\partial_\tau\,A_-(\tau) }
{(1-k A_+ A_-)^2}}\right],\\
& A_+(\zeta)=-\frac 1k \exp[-8\varepsilon_0 k \zeta],\\
& A_-(\tau)=\exp\left[{2\varepsilon_0 \int
\frac{d\tau}{m^2(\tau)+1}}\right],\\
&\eta=2(\partial_\tau m -m \partial_\tau\rho).
\end{eqnarray}

The original fields $\Omega_{a,b}$ then  read:
\begin{eqnarray}\label{fields1}
&\Omega_a=\frac{2 \varepsilon_0}{\sqrt{m^2(\tau)+1}}
\mathrm{sech}(\varphi),\\
& \Omega_b=\frac{2 \varepsilon_0\,m(\tau)}{m^2(\tau)+1}
\tanh(\varphi)+\frac{1}{2}
\frac{\partial_\tau m(\tau)}{m^2(\tau)+1},\\
&\varphi=-4k\,\varepsilon_0\,\zeta+\int \frac{ \,\varepsilon_0\,
d\tau}{m^2(\tau)+1},
\end{eqnarray}
where $\varepsilon_0$ is a real arbitrary constant defining the
amplitude of slow-light soliton. The background field $\Omega(\tau)$
reads:
\begin{equation}\label{Omega_tau}
    \Omega(\tau)=
\frac{\frac12\partial_\tau m(\tau)-2 \varepsilon_0\,m(\tau)}
{m^2(\tau)+1}.
\end{equation}

The function $\Omega(\tau)$ describes the controlling field, which
governs the dynamics of the system. The time dependence of this
function is determined by modulation of the intensity of the
auxiliary laser. As can be readily seen, the velocity of the
slow-light soliton reads
\begin{equation}\label{velo}
    v_g=\frac{1}{4k}\frac{1}{m^2(\tau)+1}.
\end{equation}
For a constant controlling field $\Omega(\tau)=\Omega_0$, and in the
simplifying approximation
$\frac{\Omega_0^2}{\varepsilon_0^2}<\!\!<1$, the group velocity of
the slow-light soliton conform to the result of linear theory:
\begin{equation}\label{ss_vg}
    v_g\approx
c\frac{\Omega_0^2}{2\nu_0}.
\end{equation}
Expression Eq.~(\ref{ss_vg}) immediately suggests that the signal
stops, when $\Omega_0=0$. Therefore, this  expression  is the main
motivational source for the works on slow-light solitons (see
\cite{Rybin:2004},\cite{ryb12} and references therein). We envisage
the following dynamics scenario. We assume that the slow-light
soliton was created in the system before the time $\tau=0$ and is
propagating on the background of the constant controlling field
$\Omega_0$. Suppose that at the moment  $\tau=0$ the laser source of
the controlling field is switched off. We assume that after this
moment the background field will  decay reasonably rapidly, as
described by a "switch-off" function $f(\tau)$. The front of the
vanishing controlling field, described by the function $f(\tau)$,
will then propagate into the medium, starting from the point
$\zeta=0$, where the laser is placed. The state of the quantum
system Eq.(\ref{init_fields0}) is dark for the controlling field.
Therefore the medium is transparent for the spreading front of the
vanishing field, which then propagates  with the speed of light,
eventually overtaking the slow-light soliton and stopping it. To
realize this scenario, we assume the controlling field
$\Omega(\tau)$ to be constant $\Omega_0$ for negative $\tau$ and a
$\tau$ dependent switch-off function $f(\tau)$ for positive $\tau$,
i.e. $\Omega(\tau)=\Omega_0 \Theta(-\tau)+f(\tau)\Theta(\tau)$. Here
$\Theta(\tau)$ is the Heaviside step function, while
$f(0)=\Omega_0$.
%~\footnote{We can cut off the tail of $\Omega(\tau)$ for $\tau
%\ge T$ provided that it is sufficiently small.}
In this setting, the distance that the soliton travels until full
stop is

\begin{equation}\label{dist}
    {\cal L}=\frac{1}{4k}\int_0^\infty \!\!\!
    \frac{d\tau}{m^2(\tau)+1}.
\end{equation}
This distance designates a geometrical point in the medium, where
the slow-light soliton   disappears, writing  itself into the medium
as a standing memory bit in the form of a localized polarization
cluster.

From Eq.(\ref{Omega_tau}) a number of exactly solvable regimes for
the stopping of the slow-light soliton can be identified. For
$m(\tau)=e^{\alpha\,\tau}$ and $\tau>0,\;\alpha>0$, we obtain
$f(\tau)=\left(\frac{\alpha}{4}-\varepsilon_0\right) \,
\mathrm{sech}(\alpha\,\tau)$ and
$${\cal L}=\frac{1}{8\alpha k}\ln2\;.$$
Hence, the slower the field decays to zero, i.e. for smaller
$\alpha$, the longer the distance that the soliton travels in the
medium is. Another physically interesting case $f(\tau)=\Omega_0
e^{-\alpha\tau}$ was discussed in \cite{ryb12}.

{\em Discussion}. In this paper we derived the slow-light soliton
from the  sufficient condition Eq.(\ref{central}). In fact, the
inverse scattering analysis as applied in \cite{ryb12} to
Eqs.(\ref{Maxwell_r}),(\ref{Schrod})
 shows that   for slow-light solitons a stronger
 condition holds, namely

\begin{equation}
\psi_3=-\frac1{2|\lambda-\Delta|}\Omega_a,\label{central_1}
\end{equation}
where   $\lambda$ is a complex number parameterizing the soliton (in
our case $\lambda=i\varepsilon_0$). This means that the condition
Eq.(\ref{central}) is also the {\it necessary} condition for the
slow-light solitons to exist with
$k=\frac{\nu_0}{8|\lambda-\Delta|^2}$.

In a forthcoming publication we will explain a fascinating analogy
between the stopping of a slow-light soliton and the formation of a
black hole in 2D gravity.

\bibliography{SL_revisited_ext}

\begin{thebibliography}{27}
\expandafter\ifx\csname natexlab\endcsname\relax\def\natexlab#1{#1}\fi
\expandafter\ifx\csname bibnamefont\endcsname\relax
  \def\bibnamefont#1{#1}\fi
\expandafter\ifx\csname bibfnamefont\endcsname\relax
  \def\bibfnamefont#1{#1}\fi
\expandafter\ifx\csname citenamefont\endcsname\relax
  \def\citenamefont#1{#1}\fi
\expandafter\ifx\csname url\endcsname\relax
  \def\url#1{\texttt{#1}}\fi
\expandafter\ifx\csname urlprefix\endcsname\relax\def\urlprefix{URL }\fi
\providecommand{\bibinfo}[2]{#2}
\providecommand{\eprint}[2][]{\url{#2}}

\bibitem[{\citenamefont{Hau et~al.}(1999)\citenamefont{Hau, Harris, Dutton, and
  Behroozi}}]{Hau:1999}
\bibinfo{author}{\bibfnamefont{L.~N.} \bibnamefont{Hau}},
  \bibinfo{author}{\bibfnamefont{S.~E.} \bibnamefont{Harris}},
  \bibinfo{author}{\bibfnamefont{Z.}~\bibnamefont{Dutton}}, \bibnamefont{and}
  \bibinfo{author}{\bibfnamefont{C.~H.} \bibnamefont{Behroozi}},
  \bibinfo{journal}{Lett.\ to Nature} \textbf{\bibinfo{volume}{397}},
  \bibinfo{pages}{594} (\bibinfo{year}{1999}).

\bibitem[{\citenamefont{Liu et~al.}(2001)\citenamefont{Liu, Dutton, Behroozi,
  and Hau}}]{Liu:2001}
\bibinfo{author}{\bibfnamefont{C.}~\bibnamefont{Liu}},
  \bibinfo{author}{\bibfnamefont{Z.}~\bibnamefont{Dutton}},
  \bibinfo{author}{\bibfnamefont{C.~H.} \bibnamefont{Behroozi}},
  \bibnamefont{and} \bibinfo{author}{\bibfnamefont{L.~V.} \bibnamefont{Hau}},
  \bibinfo{journal}{Lett.\ to Nature} \textbf{\bibinfo{volume}{409}},
  \bibinfo{pages}{490} (\bibinfo{year}{2001}).

\bibitem[{\citenamefont{Phillips et~al.}(2001)\citenamefont{Phillips,
  Fleischhauer, Mair, Walsworth, and Lukin}}]{Phillips:2001}
\bibinfo{author}{\bibfnamefont{D.~F.} \bibnamefont{Phillips}},
  \bibinfo{author}{\bibfnamefont{A.}~\bibnamefont{Fleischhauer}},
  \bibinfo{author}{\bibfnamefont{A.}~\bibnamefont{Mair}},
  \bibinfo{author}{\bibfnamefont{R.~L.} \bibnamefont{Walsworth}},
  \bibnamefont{and} \bibinfo{author}{\bibfnamefont{M.~D.} \bibnamefont{Lukin}},
  \bibinfo{journal}{Phys. Rev. Lett.} \textbf{\bibinfo{volume}{86}},
  \bibinfo{pages}{783} (\bibinfo{year}{2001}).

\bibitem[{\citenamefont{Bajcsy et~al.}(2003)\citenamefont{Bajcsy, Zibrov, and
  Lukin}}]{Bajcsy:2003}
\bibinfo{author}{\bibfnamefont{M.}~\bibnamefont{Bajcsy}},
  \bibinfo{author}{\bibfnamefont{A.~S.} \bibnamefont{Zibrov}},
  \bibnamefont{and} \bibinfo{author}{\bibfnamefont{M.~D.} \bibnamefont{Lukin}},
  \bibinfo{journal}{Lett.\ to Nature} \textbf{\bibinfo{volume}{426}},
  \bibinfo{pages}{638} (\bibinfo{year}{2003}).

\bibitem[{\citenamefont{Braje et~al.}(2003)\citenamefont{Braje, Balic, Yin, and
  Harris}}]{Braje:2003}
\bibinfo{author}{\bibfnamefont{D.~A.} \bibnamefont{Braje}},
  \bibinfo{author}{\bibfnamefont{V.}~\bibnamefont{Balic}},
  \bibinfo{author}{\bibfnamefont{G.~Y.} \bibnamefont{Yin}}, \bibnamefont{and}
  \bibinfo{author}{\bibfnamefont{S.~E.} \bibnamefont{Harris}},
  \bibinfo{journal}{Phys. Rev. A} \textbf{\bibinfo{volume}{68}},
  \bibinfo{pages}{041801(R)} (\bibinfo{year}{2003}).

\bibitem[{\citenamefont{Mikhailov et~al.}(2004)\citenamefont{Mikhailov,
  Sautenkov, Rostovtsev, and Welch}}]{Mikhailov:2004}
\bibinfo{author}{\bibfnamefont{E.~E.} \bibnamefont{Mikhailov}},
  \bibinfo{author}{\bibfnamefont{V.~A.} \bibnamefont{Sautenkov}},
  \bibinfo{author}{\bibfnamefont{Y.~V.} \bibnamefont{Rostovtsev}},
  \bibnamefont{and} \bibinfo{author}{\bibfnamefont{G.~R.} \bibnamefont{Welch}},
  \bibinfo{journal}{J. Opt. Soc. Am. B} \textbf{\bibinfo{volume}{21}},
  \bibinfo{pages}{425} (\bibinfo{year}{2004}).

\bibitem[{\citenamefont{Turukhin et~al.}(2002)\citenamefont{Turukhin,
  Sudarshanam, Shahriar, Musser, S.Ham, and Hemmer}}]{Turukhin:2002}
\bibinfo{author}{\bibfnamefont{A.~V.} \bibnamefont{Turukhin}},
  \bibinfo{author}{\bibfnamefont{V.~S.} \bibnamefont{Sudarshanam}},
  \bibinfo{author}{\bibfnamefont{M.~S.} \bibnamefont{Shahriar}},
  \bibinfo{author}{\bibfnamefont{J.~A.} \bibnamefont{Musser}},
  \bibinfo{author}{\bibfnamefont{B.}~\bibnamefont{S.Ham}}, \bibnamefont{and}
  \bibinfo{author}{\bibfnamefont{P.~R.} \bibnamefont{Hemmer}},
  \bibinfo{journal}{Phys.\ Rev.\ Lett.} \textbf{\bibinfo{volume}{88}},
  \bibinfo{pages}{023602} (\bibinfo{year}{2002}).

\bibitem[{\citenamefont{Bigelow et~al.}(2003)\citenamefont{Bigelow, Lepeshkin,
  and Boyd}}]{Bigelow:2003}
\bibinfo{author}{\bibfnamefont{M.~S.} \bibnamefont{Bigelow}},
  \bibinfo{author}{\bibfnamefont{N.~N.} \bibnamefont{Lepeshkin}},
  \bibnamefont{and} \bibinfo{author}{\bibfnamefont{R.~W.} \bibnamefont{Boyd}},
  \bibinfo{journal}{Science} \textbf{\bibinfo{volume}{301}},
  \bibinfo{pages}{200} (\bibinfo{year}{2003}).

\bibitem[{\citenamefont{Soljacic and Joannopoulos}(2004)}]{Soljacic:2004}
\bibinfo{author}{\bibfnamefont{M.}~\bibnamefont{Soljacic}} \bibnamefont{and}
  \bibinfo{author}{\bibfnamefont{J.~D.} \bibnamefont{Joannopoulos}},
  \bibinfo{journal}{Nature Materials} \textbf{\bibinfo{volume}{3}},
  \bibinfo{pages}{213} (\bibinfo{year}{2004}).

\bibitem[{\citenamefont{Kocharovskaya et~al.}(2001)\citenamefont{Kocharovskaya,
  Rostovtsev, and Scully}}]{Kocharovskaya:2001}
\bibinfo{author}{\bibfnamefont{O.}~\bibnamefont{Kocharovskaya}},
  \bibinfo{author}{\bibfnamefont{Y.}~\bibnamefont{Rostovtsev}},
  \bibnamefont{and} \bibinfo{author}{\bibfnamefont{M.~O.}
  \bibnamefont{Scully}}, \bibinfo{journal}{Phys. Rev. Lett.}
  \textbf{\bibinfo{volume}{86}}, \bibinfo{pages}{628} (\bibinfo{year}{2001}).

\bibitem[{\citenamefont{Dutton and Hau}(2004)}]{Dutton:2004}
\bibinfo{author}{\bibfnamefont{Z.}~\bibnamefont{Dutton}} \bibnamefont{and}
  \bibinfo{author}{\bibfnamefont{L.~V.} \bibnamefont{Hau}},
  \bibinfo{journal}{Phys. Rev. A} \textbf{\bibinfo{volume}{70}},
  \bibinfo{pages}{053831} (\bibinfo{year}{2004}).

\bibitem[{\citenamefont{Harris}(1997)}]{Harris:1997}
\bibinfo{author}{\bibfnamefont{S.~E.} \bibnamefont{Harris}},
  \bibinfo{journal}{Phys. Today} \textbf{\bibinfo{volume}{50(7)}},
  \bibinfo{pages}{36} (\bibinfo{year}{1997}).

\bibitem[{\citenamefont{Lukin}(2003)}]{Lukin:2003}
\bibinfo{author}{\bibfnamefont{M.~D.} \bibnamefont{Lukin}},
  \bibinfo{journal}{Rev. Mod. Phys.} \textbf{\bibinfo{volume}{75}},
  \bibinfo{pages}{457} (\bibinfo{year}{2003}).

\bibitem[{\citenamefont{Grobe et~al.}(1994)\citenamefont{Grobe, Hioe, and
  Eberly}}]{Grobe:1994}
\bibinfo{author}{\bibfnamefont{R.}~\bibnamefont{Grobe}},
  \bibinfo{author}{\bibfnamefont{F.~T.} \bibnamefont{Hioe}}, \bibnamefont{and}
  \bibinfo{author}{\bibfnamefont{J.~H.} \bibnamefont{Eberly}},
  \bibinfo{journal}{Phys.\ Rev.\ Lett.} \textbf{\bibinfo{volume}{73}},
  \bibinfo{pages}{3183} (\bibinfo{year}{1994}).

\bibitem[{\citenamefont{Eberly}(1995)}]{Eberly:1995}
\bibinfo{author}{\bibfnamefont{J.~H.} \bibnamefont{Eberly}},
  \bibinfo{journal}{Quant. Semiclass. Opt.} \textbf{\bibinfo{volume}{7}},
  \bibinfo{pages}{373} (\bibinfo{year}{1995}).

\bibitem[{\citenamefont{Andreev}(1998)}]{andreev:1998}
\bibinfo{author}{\bibfnamefont{A.~V.} \bibnamefont{Andreev}},
  \bibinfo{journal}{JETP} \textbf{\bibinfo{volume}{86}}, \bibinfo{pages}{412}
  (\bibinfo{year}{1998}).

\bibitem[{\citenamefont{Dey and Agarwal}(2003)}]{Dey:2003}
\bibinfo{author}{\bibfnamefont{T.~N.} \bibnamefont{Dey}} \bibnamefont{and}
  \bibinfo{author}{\bibfnamefont{G.~S.} \bibnamefont{Agarwal}},
  \bibinfo{journal}{Phys. Rev. A} \textbf{\bibinfo{volume}{67}},
  \bibinfo{pages}{033813} (\bibinfo{year}{2003}).

\bibitem[{\citenamefont{Kozlov and Eberly}(2000)}]{Kozlov:2000}
\bibinfo{author}{\bibfnamefont{V.~V.} \bibnamefont{Kozlov}} \bibnamefont{and}
  \bibinfo{author}{\bibfnamefont{J.~H.} \bibnamefont{Eberly}},
  \bibinfo{journal}{Opt. Commun.} \textbf{\bibinfo{volume}{179}},
  \bibinfo{pages}{85} (\bibinfo{year}{2000}).

\bibitem[{\citenamefont{Dutton et~al.}(2001)\citenamefont{Dutton, Budde, Slowe,
  and Hau}}]{Dutton:2001}
\bibinfo{author}{\bibfnamefont{Z.}~\bibnamefont{Dutton}},
  \bibinfo{author}{\bibfnamefont{M.}~\bibnamefont{Budde}},
  \bibinfo{author}{\bibfnamefont{C.}~\bibnamefont{Slowe}}, \bibnamefont{and}
  \bibinfo{author}{\bibfnamefont{L.~V.} \bibnamefont{Hau}},
  \bibinfo{journal}{Science} \textbf{\bibinfo{volume}{293}},
  \bibinfo{pages}{663} (\bibinfo{year}{2001}).

\bibitem[{\citenamefont{Matsko et~al.}(2001)\citenamefont{Matsko, Rostovtsev,
  Kocharovskaya, Zibrov, and Scully}}]{Matsko:2001}
\bibinfo{author}{\bibfnamefont{A.~B.} \bibnamefont{Matsko}},
  \bibinfo{author}{\bibfnamefont{Y.~V.} \bibnamefont{Rostovtsev}},
  \bibinfo{author}{\bibfnamefont{O.}~\bibnamefont{Kocharovskaya}},
  \bibinfo{author}{\bibfnamefont{A.~S.} \bibnamefont{Zibrov}},
  \bibnamefont{and} \bibinfo{author}{\bibfnamefont{M.~O.}
  \bibnamefont{Scully}}, \bibinfo{journal}{Phys. Rev. A}
  \textbf{\bibinfo{volume}{64}}, \bibinfo{pages}{043809}
  (\bibinfo{year}{2001}).

\bibitem[{\citenamefont{Faddeev and Takhtadjan}(1987)}]{fad}
\bibinfo{author}{\bibfnamefont{L.~D.} \bibnamefont{Faddeev}} \bibnamefont{and}
  \bibinfo{author}{\bibfnamefont{L.~A.} \bibnamefont{Takhtadjan}},
  \emph{\bibinfo{title}{Hamiltonian Methods in the Theory of Solitons}}
  (\bibinfo{publisher}{Springer, Berlin}, \bibinfo{year}{1987}).

\bibitem[{\citenamefont{Park and Shin}(1998)}]{Park:1998}
\bibinfo{author}{\bibfnamefont{Q.~H.} \bibnamefont{Park}} \bibnamefont{and}
  \bibinfo{author}{\bibfnamefont{H.~J.} \bibnamefont{Shin}},
  \bibinfo{journal}{Phys.\ Rev.\ A} \textbf{\bibinfo{volume}{57}},
  \bibinfo{pages}{4643} (\bibinfo{year}{1998}).

\bibitem[{\citenamefont{Byrne et~al.}(2003)\citenamefont{Byrne, Gabitov, and
  Kova\v{c}i\v{c}}}]{gab}
\bibinfo{author}{\bibfnamefont{J.~A.} \bibnamefont{Byrne}},
  \bibinfo{author}{\bibfnamefont{I.~R.} \bibnamefont{Gabitov}},
  \bibnamefont{and}
  \bibinfo{author}{\bibfnamefont{G.}~\bibnamefont{Kova\v{c}i\v{c}}},
  \bibinfo{journal}{Physica D} \textbf{\bibinfo{volume}{186}},
  \bibinfo{pages}{69} (\bibinfo{year}{2003}).

\bibitem[{\citenamefont{Rybin and Vadeiko}(2004)}]{Rybin:2004}
\bibinfo{author}{\bibfnamefont{A.~V.} \bibnamefont{Rybin}} \bibnamefont{and}
  \bibinfo{author}{\bibfnamefont{I.~P.} \bibnamefont{Vadeiko}},
  \bibinfo{journal}{Journal of Optics B: Quantum and Semiclassical Optics}
  \textbf{\bibinfo{volume}{6}}, \bibinfo{pages}{416} (\bibinfo{year}{2004}).

\bibitem[{\citenamefont{Fleischhauer and Lukin}(2000)}]{Fleischhauer:2000}
\bibinfo{author}{\bibfnamefont{M.}~\bibnamefont{Fleischhauer}}
  \bibnamefont{and} \bibinfo{author}{\bibfnamefont{M.~D.} \bibnamefont{Lukin}},
  \bibinfo{journal}{Phys.\ Rev.\ Lett.} \textbf{\bibinfo{volume}{84}},
  \bibinfo{pages}{5094} (\bibinfo{year}{2000}).

\bibitem[{\citenamefont{Rybin et~al.}(2005)\citenamefont{Rybin, Vadeiko, and
  Bishop}}]{ryb12}
\bibinfo{author}{\bibfnamefont{A.~V.} \bibnamefont{Rybin}},
  \bibinfo{author}{\bibfnamefont{I.~P.} \bibnamefont{Vadeiko}},
  \bibnamefont{and} \bibinfo{author}{\bibfnamefont{A.~R.}
  \bibnamefont{Bishop}}, \bibinfo{journal}{Phys. Rev. E}
  \textbf{\bibinfo{volume}{72}}, \bibinfo{pages}{026613}
  (\bibinfo{year}{2005}).

\bibitem[{\citenamefont{Giacomini and Pinamonti}(2003)}]{Giacomini:2003}
\bibinfo{author}{\bibfnamefont{A.}~\bibnamefont{Giacomini}} \bibnamefont{and}
  \bibinfo{author}{\bibfnamefont{N.}~\bibnamefont{Pinamonti}},
  \bibinfo{journal}{J. High Energy Phys.} \textbf{\bibinfo{volume}{02}},
  \bibinfo{pages}{014} (\bibinfo{year}{2003}).

\end{thebibliography}
\end{document}